\documentclass[%
superscriptaddress,
 amsmath,amssymb,
prb,
floatfix,
twocolumn]{revtex4-1}

\usepackage{graphicx}
\usepackage{dcolumn}
\usepackage{bm}
\usepackage{libertine}

\begin{document}

\title{\sc  \LARGE Evolutionary algorithms converge towards 
evolved biological photonic structures}

\author{Mamadou Aliou Barry}
\affiliation{Universit\'e Clermont Auvergne, CNRS, Institut Pascal, 63000 Clermont-Ferrand, France}
\author{Vincent Berthier}
\affiliation{TAO, Inria, LRI, Universit\'e Paris-Saclay CNRS UMR 6823}
\author{Bodo~D.~Wilts}
\affiliation{Adolphe Merkle Institute, University of Fribourg, Chemin des Verdiers 4, 1700 Fribourg, Switzerland}
\author{Marie-Claire Cambourieux}
\author{R\'emi Poll\`es}
\affiliation{Universit\'e Clermont Auvergne, CNRS, Institut Pascal, 63000 Clermont-Ferrand, France}
\author{Olivier Teytaud}
\affiliation{TAO, Inria, LRI, Universit\'e Paris-Saclay CNRS UMR 6823}
\affiliation{Google Brain Z\"urich, Brandschenkestrasse 110,
8002 Z\"urich}
\author{Emmanuel Centeno}
\affiliation{Universit\'e Clermont Auvergne, CNRS, Institut Pascal, 63000 Clermont-Ferrand, France}
\author{Nicolas Biais}
\affiliation{Graduate Center of CUNY and Department of Biology, CUNY Brooklyn College, NY 11210 New York, U.S.A.}
\author{Antoine Moreau$^*$}
\affiliation{Universit\'e Clermont Auvergne, CNRS, Institut Pascal, 63000 Clermont-Ferrand, France}
\email{antoine.moreau@uca.fr}

\date{\today}

\begin{abstract}
 Nature features a plethora of extraordinary photonic architectures that have been optimized through natural evolution. While numerical optimization is increasingly and successfully used in photonics, it has yet to replicate any of these complex naturally occurring structures. Using evolutionary algorithms directly inspired by natural evolution, we have retrieved emblematic natural photonic structures, indicating how such regular structures might have spontaneously emerged in nature and to which precise optical or fabrication constraints they respond. Comparisons between algorithms show that recombination between individuals inspired by sexual reproduction confers a clear advantage in this context of modular problems and suggest further ways to improve the algorithms. Such an {\em in silico} evolution can also suggest original and elegant solutions to practical problems, as illustrated by the design of counter-intuitive anti-reflective coating for solar cells.
\end{abstract}

\maketitle

Nature features a plethora of photonic architectures producing the most vivid optical effects\cite{vukusic2003photonic,kinoshita2008physics}. These ubiquitous structures have been optimized through natural evolution during millions of years and include natural photonic crystals\cite{parker2001} as well as the extravagant architectures that can be found on {\em Morpho} butterfly wings. With the development of fast simulation tools for optics\cite{lalanne1996highly,granet1996efficient}, numerical optimization has been increasingly used  in photonics\cite{martin1995synthesis,yang2001evolutionary,gondarenko2006spontaneous}, recently producing designs with increased performances\cite{piggott2015inverse,shen2015integrated,bruck2016all}. However, these structures usually do not possess the regularity and elegance of the natural ones.
This is made paradoxical by the fact that for these particularly complex optimization problems, evolutionary algorithms\cite{de,pso,cma,rechenberg73} {\em i.e.} optimization methods inspired by evolutionary strategies, have been repeatedly tested. We are left to wonder whether natural structures are regular or periodical because of such architectures are easier to fabricate, or whether such a regularity is imposed for optical reasons.

In the domain of thin film optical filters for instance, optimization has been used extensively to help design complex optical filters, like the ones at the core of the multiplexed internet traffic. But all the algorithms, including specific methods like the needle method\cite{larouche}, can be considered to be local optimization methods, starting with a solution which is already satisfactory and improving it. In photonics in general, researchers feel that (i) all the existing optimization techniques have already been thoroughly tested and (ii) specific methods that are adapted to the photonic problem. This state of affairs is unfortunate as optimization techniques have immensely improved in the last ten years, particularly evolutionary algorithms which are versatile global optimization techniques.

Evolutionary algorithms\cite{nm,de,pso,cma,rechenberg73} are computational trial-and-error algorithms that aim at finding optimal solutions to well posed mathematical problems while being inspired by evolutionary processes. In general, evolutionary algorithms consider a population of individuals, where each individual corresponds to a potential solution. An objective function allows to rate the fitness of a solution/individual: the lower the objective function, the better the solution and thus the more "fit" the corresponding individual. The population then evolves: while individuals that are not fit enough are "eliminated", new individuals are subsequently created, for example by combining the characteristics of better individuals. Through this {\em in silico} evolution, the average fitness of the individuals increases and ultimately leads to the ``best'' possible solution to the problem posed.

Here we show, by applying state of the art evolutionary algorithms to increasingly complex optical problems, that solutions corresponding exactly to natural photonic architectures can be retrieved. We establish a clear link between the constraints we impose and the structures that emerge as a result, which allows to understand the precise role of each feature. Then, we compare the different algorithms in order to assess which evolutionary strategies are the most efficient. Finally we show how this method can be applied to key problems in the optical sciences, e.g. enhancing the absorption of light in solar cells.

\begin{figure*}[!htbp]
\centering
 {\includegraphics[width=0.9\linewidth]{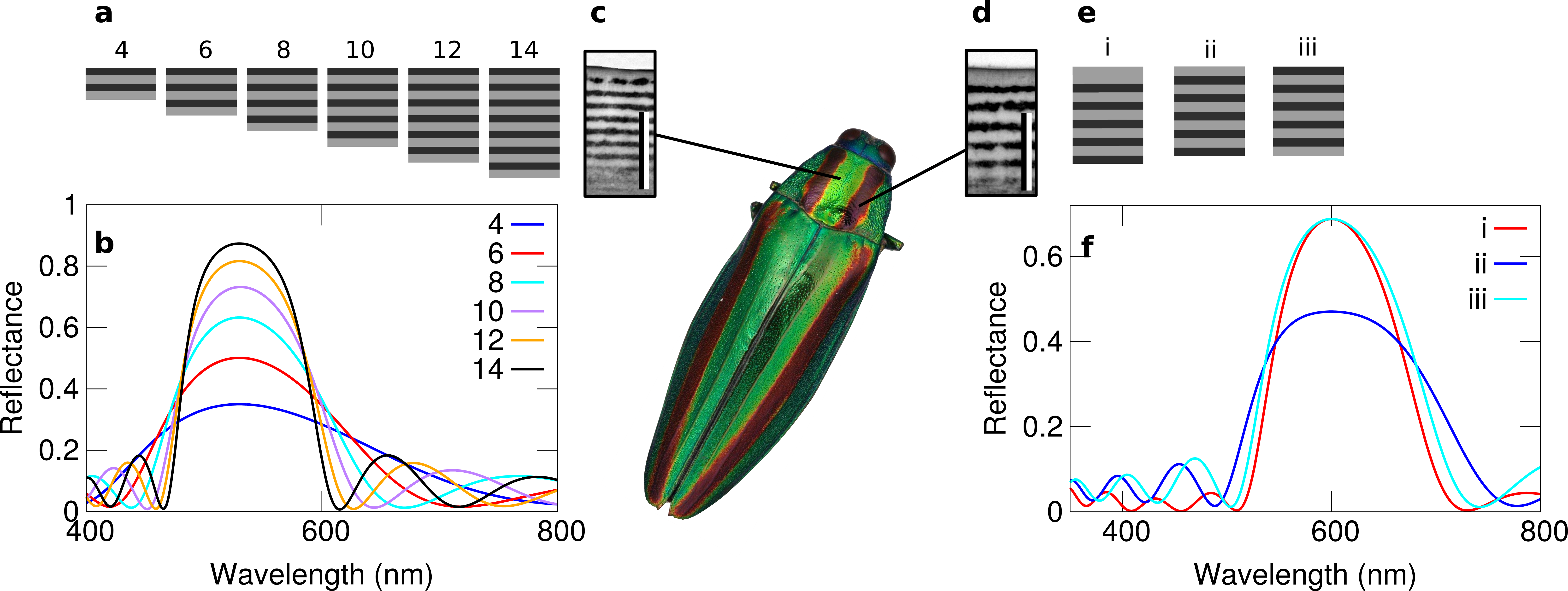}}
\caption{\textbf{Retrieving dielectric mirrors through optimization.} \textbf{a} Solutions obtained through optimization for different numbers of layers, ranging from 4 to 14. \textbf{b} Associated reflectance spectrum. \textbf{c,d} TEM images of the cuticular surface structure of the Japanese Jewel beetle, {\em Chrysochroa fulgidissima} \cite{schenk2013japanese} (green and purple part of the elytron, respectively), bar: 1µm. \textbf{e} Dielectric mirrors beginning with the lower index with a $\lambda/2$ (i) and $\lambda/4$ (ii) thickness and dielectric mirror beginning with the higher index for the same number of layers (iii). \textbf{f} Corresponding reflectance spectra showing similar efficiencies for structures (i) and (iii).}
\label{fig:bragg}
\end{figure*}

{\bf Emergence of Bragg mirrors}. Light can be reflected by Bragg mirrors (periodic multilayered structures composed of transparent materials with alternating refractive index (RI)\cite{macleod2001thin}) more efficiently than by any metallic mirror. Such multilayered structures constitute the simplest example of photonic crystals, structures owing their optical properties to their regularity. Not surprisingly, such dielectric-only mirrors can be found in nature on the integument of animals (see Fig. \ref{fig:bragg}) and have also been widely employed in technical applications. This suggests that these multilayered optical devices are somehow optimal, but paradoxically, despite the apparent simplicity of the problem, no computational optimization algorithm has ever yielded Bragg mirrors as a solution, including early attempts using evolutionary algorithms\cite{martin1995synthesis}.

 In this work, we have employed five different, state-of-the-art  optimization algorithms\cite{nm,de,pso,cma,rechenberg73} best suited for our setting, {\em i.e.} complex real-world problems for which little is known {\em a priori}. The algorithms differ in their original inspiration (see Supplementary Information). The first (1+1 - ES) is inspired by the evolution of bacteria, with local mutations taking a central place\cite{rechenberg73}; the second (Differential Evolution\cite{de}) is inspired by the evolution process of sexual selection and includes recombination between successful individuals; the last ones are less related to evolution. Particle Swarm Optimization\cite{pso} is inspired by the behavior of swarms, while Covariance Matrix Adaptation and Nelder-Mead\cite{nm} are more artificial algorithms based on profound mathematical considerations.

 We first begin by considering multilayers of transparent materials, the simplest of all photonic structures, to investigate whether the algorithms are able to produce regular structures as optimal solutions -- having dielectric mirrors in mind.
The objective function is defined so that the algorithms simply maximize the reflection coefficient of the structure for a given wavelength of the incident light -- and computed using a freely available simulation tool for electromagnetic optics\cite{defrance16}. The algorithms are free to modify the thicknesses and the RI of the individual layers. For several numbers of layers ranging from 4 to 40, we run up to 100 optimizations for each algorithm. The only constraint we impose is that the RI has to be in the range of 1.4 to 1.7,  typical for organic materials \cite{leertouwer2011}. The best solution of our optimization scheme is consistently a stack of alternating layers with RI of 1.7 and 1.4, respectively, with a thickness of a quarter of the wavelength (see Supplementary Information) -- which corresponds exactly to the Bragg mirror. We underline here that it is almost impossible to assess whether an optimization has found the actual optimal solution. However, the consistent correspondence of simulated and natural structures makes us confident we actually have found an optimum. Moreover, this result shows that  the optical constraint of reflecting light efficiently alone explains the emergence of a regular pattern -- which hints as to why dielectric mirrors are so ubiquitous in nature.

{\bf Better understanding natural designs}. When carried out with a higher number (up to 40) of layers, the optimization scheme produces more chaotic structures (see Supplementary Information), still systematically indicative of the fact that the RI should only alternate between 1.4 and 1.7, the most extreme values allowed - and start with the higher index facing the outermost \textit{in vacuo} layer. Dielectric mirrors in nature however consist of higher index layers, predominantly made of melanin, comprised in a matrix of cuticular chitin and these usually begin with chitin, presenting the lowest RI. 

When we force the algorithms to use this extreme values of the refractive index and to begin with a lower RI, a slightly different dielectric mirror emerges, with a first layer twice as thick ($\lambda$/2) as the other low RI layers. A physical analysis allows to understand the functional role of this layer. A lower RI layer with quarter-wave thickness and a RI of 1.4 in fact constitutes an anti-reflective coating - thus lowering the reflectance.  A first layer of half a wavelength  constitutes actually an ``absent layer''\cite{macleod2001thin} allowing to obtain the performances of a standard dielectric mirror while beginning with a lower refractive index. Man-made dielectric mirrors always begin with the higher RI medium for the above reasons\cite{macleod2001thin}. However, existing literature shows that part of the elytral cuticles of the Japanese jewel beetle,  \textit{Chrysochroa fulgidissima} (see Fig.~\ref{fig:bragg}), are actually covered with multilayers following this exact design principle\cite{schenk2013japanese}.

For a low number of layers, like in the case of the purple stripes of \textit{Chrysochroa fulgidissima}, the reflectance can be doubled with this simple recipe (see Fig. \ref{fig:bragg}). This indicates that beginning with a low index layer is a fabrication constraint and that the design we have found by optimization is the solution that has emerged spontaneously in this species as a response.

\begin{figure*}[ht]
\centering
 {\includegraphics[width=0.9\textwidth]{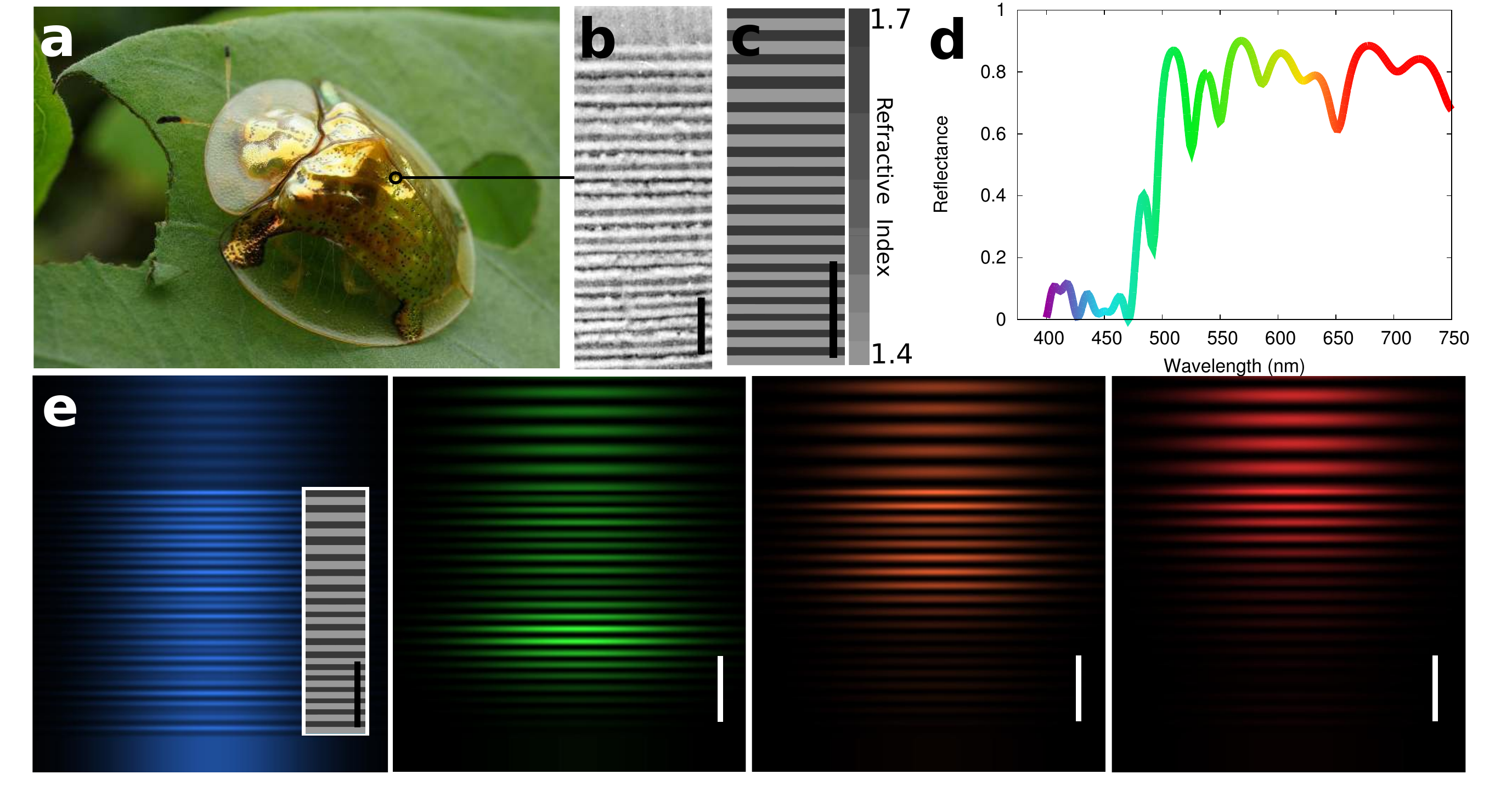}}
\caption{{\bf Retrieving chirped dielectric mirrors}. \textbf{a} {\em Aspidomorpha tecta}, "Fool's gold beetle", photograph by Indri Basuki. \textbf{b} TEM image of the structure on the cuticule, taken from \cite{neville1977metallic}. \textbf{c} Result of the optimization by evolutionary algorithms with a larger period at the top than at the bottom. \textbf{d} Reflection spectrum of the structure. \textbf{e} Electric field distribution map upon normal-incidence illumination showing how different wavelength (and thus colors) are reflected at different depth in the chirped dielectric mirror. From left to right: blue (400 nm), green (530 nm), orange (600 nm) and red (700 nm). Scale bars: 1 $\mu$m.
}
\label{fig:chirped}
\end{figure*}

{\bf Retrieving chirped dielectric mirrors.} Numerical optimization of photonic structures is thus able to point out specific features of natural structures that went previously unnoticed as well as to retrieve perfectly regular optimal structures. As shown in Fig. \ref{fig:bragg}, periodic dielectric mirrors however possess a fixed bandwidth, determined solely by the RI contrast and the number of layers. Trying to tackle a more complex problem for which solutions exist both in nature and in technology, we subsequently changed the objective function to search for a broadband dielectric mirror,\textit{i.e.} a multilayered structure that reflects several wavelengths in an interval much larger than the bandwidth of a dielectric mirror. The most favorable designs produced by the algorithms are dielectric mirrors with slowly varying thicknesses, as shown in Fig. \ref{fig:chirped}, in which the different wavelength are reflected at different depths. Such devices are  known as chirped dielectric mirrors, commonly found on metallic scarab beetles\cite{seago2009} and butterfly pupae \cite{steinbrecht1985}, but also employed in broad-band optics\cite{chirped97}. This shows that regular but non-periodic architectures, which are much more complex than Bragg mirrors, can also be retrieved by numerical optimization. 

\begin{figure*}[ht]
\centering
 {\includegraphics[width=\textwidth]{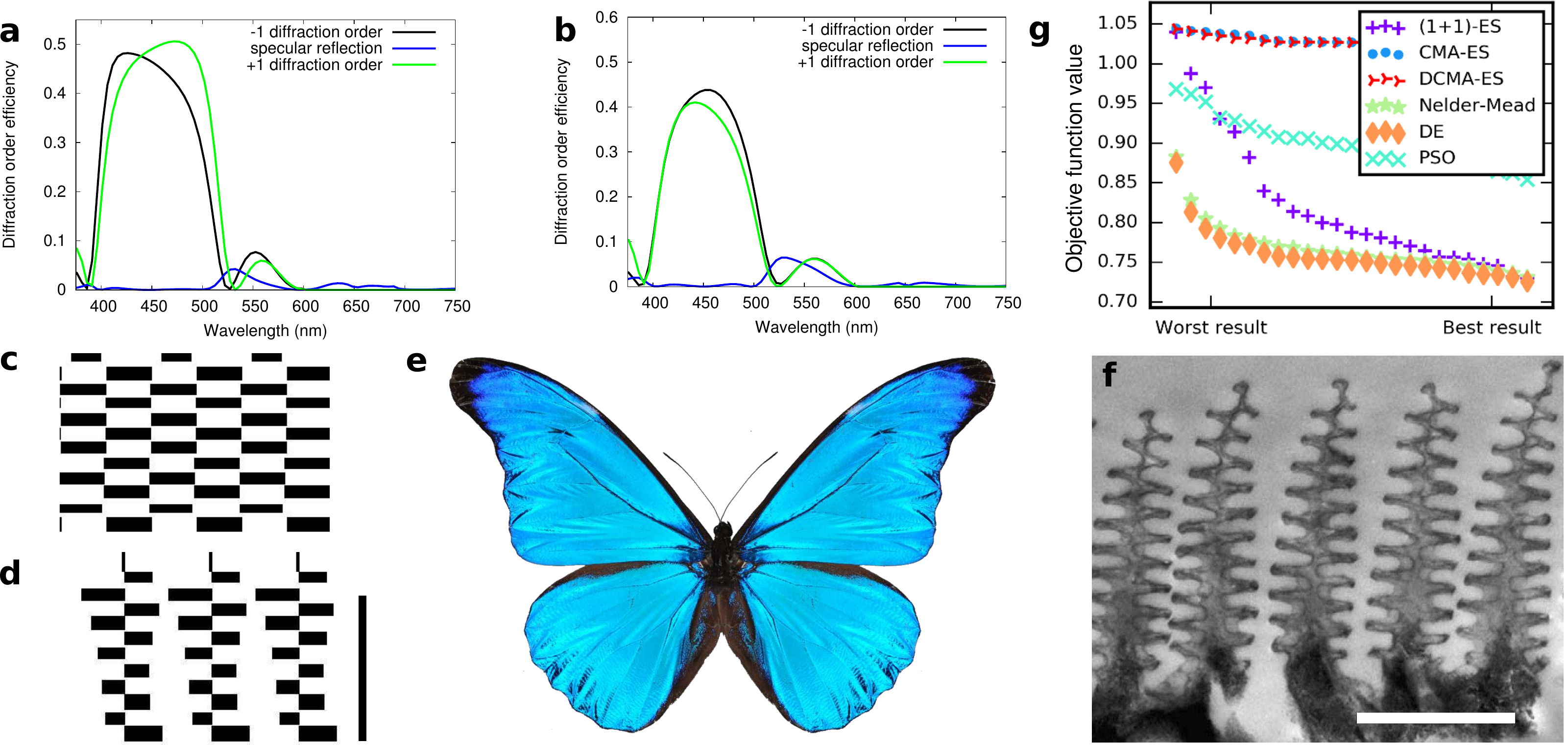}}
\caption{{\bf Retrieving the \textit{Morpho} wing scale architecture}.\textbf{a} Diffraction efficiency of the diffraction orders for the optimal structure (shown in {\bf c}) found by the algorithms with no constraint except for the horizontal periodicity (fixed). \textbf{b} Diffraction efficiencies for the structure found (shown in {\bf d} when including a fabrication constraint and a pressure towards a lighter architecture. The bar represents 1 $\mu$m. \textbf{e} Actual view of a {\em Morpho rethenor}, photograph by  John Nielsen. {\bf f} TEM images of the cuticular surface of {\em Morpho rhetenor} [taken from \cite{stavenga2011polarized}]. The bar represents 1 $\mu$m. {\bf g} Score (lowest value of the objective function reached) for each algorithm with 12 layers and  penalization, corresponding to the case illustrated in {\bf d}; the x-axis represents the different runs, sorted (best run on the right). See Supplementary Information.}
\label{fig:morpho}
\end{figure*}

{\bf Retrieving the Morpho wing scale architecture.} The most emblematic photonic structure in nature is undoubtedly the architecture that can be found on \textit{Morpho} butterflies \cite{vukusic1999,Giraldo}. In \textit{Morpho} butterflies, the wing scale ridges are folded into multilayers that resemble a Christmas-tree structure, which is continuous along the length of the wing scales (see Fig. \ref{fig:morpho}). The structure to be optimized is constituted of rectangular blocks of cuticular chitin with arbitrary dimensions and position, forming a periodic structure with a fixed horizontal period . Each block is separated from the others by an air layer of arbitrary thickness. The algorithms not only have to minimize the specular reflection at any given wavelength, but also to maximize the scattering of blue light (450 nm) in the higher diffraction orders, in order to reproduce the line-like scattering pattern observed in \textit{Morpho} butterfly wing scales \cite{Giraldo} (see Supplementary Information). Despite the jump in complexity, requiring advanced numerical methods\cite{lalanne1996highly,granet1996efficient} that are much more costly, the algorithms produce very efficient structures that have less than 0.0001\% specular reflection whereas 98\% is scattered into the diffraction orders due to an intertwined arrangement of blocks that resemble the \textit{Morpho}  wing scale nanostructure (see Fig. \ref{fig:morpho}). The interdigitation is clearly responsible for the almost total cancellation of the specular reflection, indicating that the evolutionary constraint of maximizing scattering made such architectures potentially emerge. 

Obviously, the actual \textit{Morpho}structure is not optically optimal in so far as the biological construction of the structure will come with a set of other constraints that are difficult to evaluate. We have thus added additional constraints with the aim of reproducing the natural structures: (i) the blocks should be on top of each other, and (ii) the structure should be as light as possible. When the latter constraint, controlled by a parameter in the objective function (see Supplementary Information), is strong enough,  structures that are optically only slightly sub-optimal and very close to actual \textit{Morpho} structures clearly emerge (see Fig. \ref{fig:morpho}). This definitely shows that evolutionary algorithms are not only able to yield regular, elegant and complex solutions to various optical problems, but that this process allows to understand the precise purpose of each of their features and even to quantify the balance between the optical and mechanical constraints.

{\bf Comparison of evolutionary strategies.} The performances of the different algorithms significantly differ. Whether multilayered or more complex structures are considered, one algorithm, Differential Evolution (DE) in its most widespread variant, consistently outperforms the others including non-evolutionary algorithms, as shown on Fig. \ref{fig:morpho} (see Supplementary Information for the complete results). This algorithm is the only one that is inspired by sexual evolution, recombination (by crossover rather than by averaging) between individuals playing a central role. At each generation, an individual generates an offspring that replaces its parent only if it is more "fit". A characteristic of a parent (like the thickness of a given layer, or the width of a block) has one in two chances to be transfered directly to the offspring -- this process, called the crossover, is partly inspired by gene dominance. Otherwise, this characteristic will take a value computed by "mixing" four individuals: the parent, two randomly chosen individuals and the best individual so far (see Supplementary Information). 

In more classical numerical problems used to compare optimization algorithms DE does not necessarily fare better - we attribute this discrepancy to the fact that the physical problems considered here are, without doubt, modular: each part of the structure, even if it interacts with the rest of the architecture, has a precise purpose that can be optimized partially independently. DE presents features that help in such a situation, namely combining exact copies of some variables and new variables from other individuals. 

We underline that many other problems in the optical science and related fields could benefit from such an approach, all the more so that DE is a very simple algorithm, surprisingly requiring only a few lines of code. Although its operators have been designed independently of the present work, DE seems to be a {\em pot-pourri} of the strategies of sexual evolution - selection of the fittest, gene crossover and mixing of up to four genomes, role of the best individual in the reproduction process.

\begin{figure}[h!]
\centering
\includegraphics[width=0.5\textwidth]{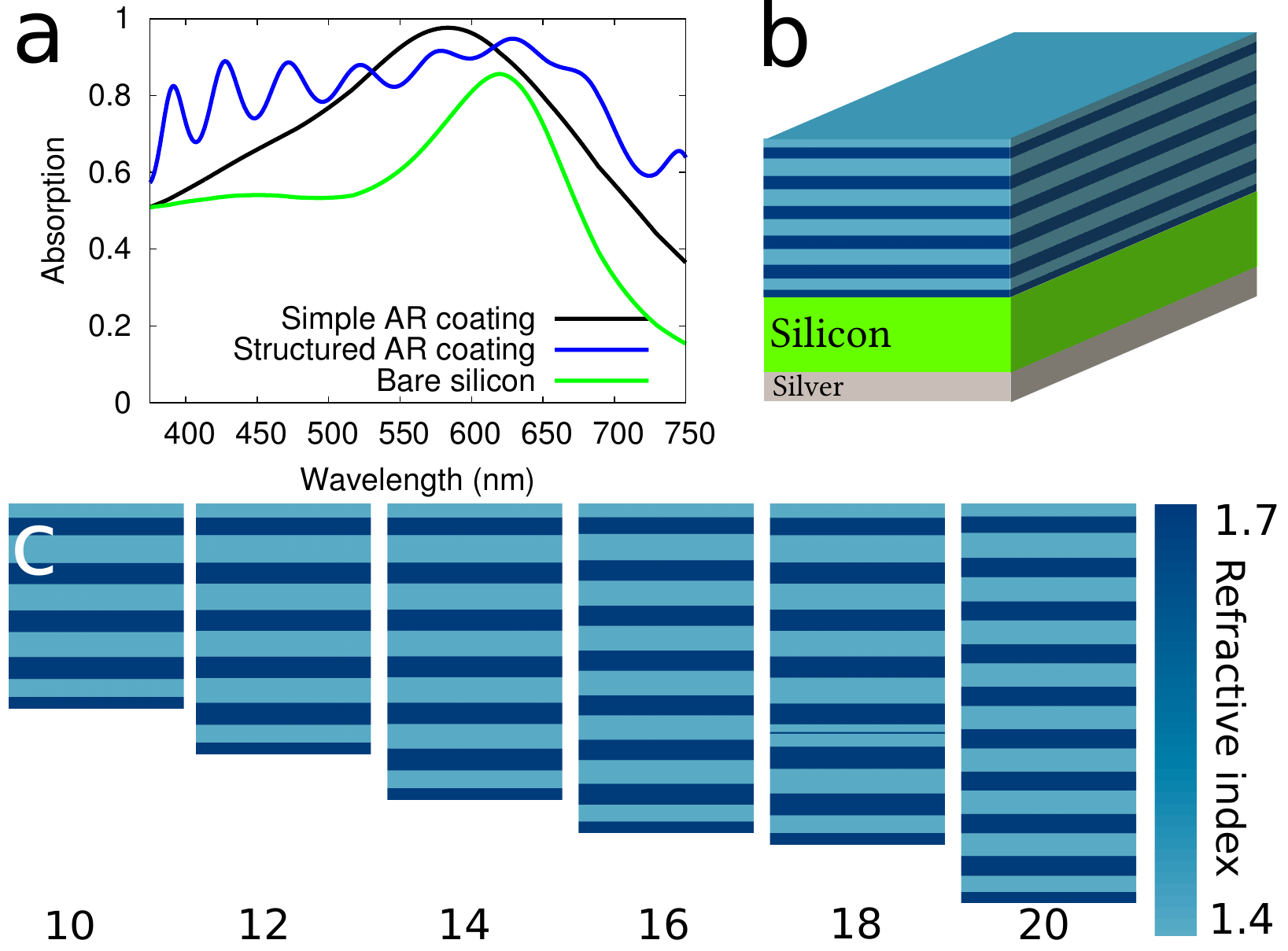}
\caption{{\bf Anti-reflective coatings produced by evolutionary optimization}. \textbf{a} Absorption spectra for a 89 nm thick amorphous silicon layer covered, bare or covered with different anti-reflective coatings.  {\bf b}, scheme of the structure with the multilayered anti-reflective coating designed by the algorithms with 12 layers. {\bf c} Results of the optimization for different numbers of layers on top of a 10 $\mu$m thick amorphous silicon layer.}
\label{fig:pv}
\end{figure}

{\bf Anti-reflective coatings produced by evolutionary optimization.} To demonstrate that the present approach is widely generalizable and not limited to natural photonic structures \textit{per se}, we  consider the problem of an anti-reflective coating on a solar cell based on amorphous silicon illuminated in normal incidence. Such coatings have been largely studied and optimized in the past for a small number of layers. Here, we run an optimization for various numbers of layers (up to 20), imposing a maximum RI contrast (alternating layers of 1.4 and 1.7 index beginning with the lower index in this case), searching for the highest solar cell performance ({\em i. e.} conversion efficiency) for two very different thicknesses of the silicon layer (89 nm and 10 $\mu$m). In both cases, the result is clearly a modified quarterwave stack reflecting only the infra-red part of the spectrum, with the two upper and two lower layers presenting a reverse pattern, different from the dielectric mirror pattern (see Fig. \ref{fig:pv} ). This result does not depend on the number of layers imposed. The performances of such a device are excellent, allowing to absorb about 80\% of the incident visible photons inside a 89 nm thick amorphous silicon layer and clearly outperforming a standard quarterwave anti-reflective coating (see Fig. \ref{fig:pv} and Supplementary Information). These multilayered coating have the unique property of reflecting infra-red light very efficiently. This could potentially prevent solar cells from overheating, which is lowering their efficiency. Our strategy allows us to conclude, very counter-intuitively, that slightly modified dielectric mirrors can be turned into efficient anti-reflective coatings.  

{\bf Concluding remarks.} Since our optimization have yield very regular structures, this means the regularity or periodicity of natural structures is the result of the optical constraints alone. In the case of the \textit{Morpho} butterfly, our results show that fabrication constraints obviously play a role, and we were even able to quantify it, but even then the vertical periodicity subsists for purely optical reasons. Our results thus suggest that optimal photonic structures should generally be expected to be periodical or regular. As already underlined above, the optimization of photonic structures has already been the subject of numerous papers, but either the periodicity was assumed \textit{a priori}\cite{cox2000band,sell2017periodic} or the resulting structures were rather disordered and very different from any naturally occurring structures\cite{piggott2015inverse,shen2015integrated,bruck2016all}, which, despite their good performances, casts doubts on the fact they are truly optimal.

While for complex enough problems, it is generally impossible to guarantee the solutions found are optimal, for photonic structures whose complexity can be arbitrarily chosen a simple procedure can be followed. The complexity of the problem must be gradually increased, keeping constant the number of evaluations of the cost function allowed for the algorithms. As the complexity grows, the structures begin to loose their regularity (see Supplementary Information). This shows the problem has become too difficult for the algorithms with such a limited budget. Paradoxically, being able to determine where this limit lies reinforces the confidence one can have in the solutions produced by the algorithms for the lowest complexity - especially if they all possess similar characteristics. Many disordered structures have probably emerged in previous works because the problem was simply too difficult for the algorithms.

 Furthermore, by showing that the various photonic structures that we have investigated are indeed optimized solutions to different well posed problems, this study provides insights regarding nature's rationale for the observed structures. Also, our results show that the designs on which technological realizations rely (e.g. dielectric mirrors) are indeed rather optimal.

The early promoters of genetic algorithms hoped they had found an ``invention machine'' able to compete with human intelligence ("human-competitive"
\cite{Koza2010}), {\em i.e.} to retrieve inventions made by humans in various domains. Using the procedure described above, we have been able to propose an original anti-reflective coating based on a Bragg mirror - making such a structure so counter-intuitive that no human could ever have come up with such a design. This suggests that modern optimization algorithms, being much more efficient than early optimization heuristics inspired by nature\cite{janikow1991experimental} can be used to compete with nature and provide a new source of inspiration.

 Improving optimization algorithms requires complex problems for which the best solution is known. This is exceedingly rare, as determining that a given solution is actually optimal can be extremely difficult, if not impossible. This work shows that nature has provided us with a with a whole class of testbed problems {\em and their solution}, which will be extremely useful to design future optimization algorithms.

In photonics, optimization problems are obviously difficult, because the optical response of even the simplest structures can be particularly complex, but they are clearly modular, as are a large part of real-world problems. This favors optimization algorithms inspired by sexual evolution strategies because they include crossovers, making them able to combine the most efficient parts of different designs.

We finally think that our results shed a new light on evolutionary processes. We emphasize however that, while evolutionary algorithms are directly inspired by evolution, they do not constitute in any way simulations of a realistic evolution.

Evolution has always been thought as a slow optimization process leading to increasingly fitter individuals through natural selection.  Our results show that such a process can actually lead to truly optimal solutions in a well defined mathematical sense. This allows to better understand why evolution produced these photonic architectures, that are so extraordinary that their spontaneous emergence seems very unlikely.

We have already stressed out that the most successful evolutionary algorithm, by far, is the only one where individuals explicitly exchange information. On particularly modular problems, it largely outperforms even Nelder-Meade, an algorithm able to take into account the global structure of the fitness landscape. By showing that these very features make the optimization process more efficient and allow to find the most complex and elegant architectures that occur in nature, our study finally suggests that sexual reproduction indeed brings an evolutionary advantage, which is still strongly debated in biology\cite{ridley,hartfield2012current}. 

\section*{Acknowledgments}

The authors would like to thank Marc Schoenauer, G\'erard Granet, Stéphane Larouche and Julien Lumeau for fruitful discussions and help, Indri Basuki for the permission to use the photograph of {\em Aspidomorpha tecta} and John Nielsen, Australia, for the permission to use the photograph of {\em Morpho rethenor}.

\bibliography{biblio}

\end{document}